\documentclass[pre,superscriptaddress,preprint,showpacs]{revtex4}

\usepackage{amsmath}
\usepackage{graphicx}
\usepackage{amssymb}

\makeatletter

\usepackage{subfigure}
\usepackage{amsfonts}

\makeatother
\begin{document}

\title{Unwinding dynamics of double-stranded polymers}

\author{M. Baiesi}
\affiliation{Department of Physics, University of Padua,  35131 Padova, Italy}
\affiliation{Institute for Theoretical Physics, K.U.Leuven,  B-3001 Leuven, Belgium}

\author{G. T. Barkema}
\affiliation{Institute for Theoretical Physics, Utrecht University, 3584CE Utrecht, The Netherlands}
\affiliation{Instituut-Lorentz, Universiteit Leiden, 2333 CA Leiden, The Netherlands}

\author{E. Carlon}
\affiliation{Institute for Theoretical Physics, K.U.Leuven,  B-3001 Leuven, Belgium}

\author{D. Panja}
\affiliation{Institute for Theoretical Physics, University of Amsterdam, Science Park 904,
1090 GL Amsterdam, The Netherlands}

\date{\today}

\begin{abstract}
We consider the unwinding of two lattice polymer strands of length
$N$ which are initially wound around each other in a double helical
conformation and evolve through Rouse dynamics. The problem relates
to quickly bringing a double-stranded polymer well above its melting
temperature, i.e., binding interactions between the strands are neglected,
and the strands separate from each other as it is entropically favorable
for them to do so. The strands unwind by rotating around each other
until they separate. We find that the process proceeds from the ends
inwards; intermediate conformations can be characterized by a tightly
wound inner part, from which loose strands are sticking out, with length
$l \sim t^{0.39}$. The total time needed for the two strands to unwind
scales as a power of $N$ as $\tau_u \sim N^{2.57\pm 0.03}$. We present a
theoretical argument which suggests that during this unwinding process,
these loose strands are far out of equilibrium.
\end{abstract}

\pacs{36.20.Ey, 
87.15.H-,     
87.14.gk      
}

\maketitle

\section{Introduction}

There are several known examples of polymers in nature that are
composed by two or more strands arranged in a helical conformation.
One is the double helical structure of DNA, in which two complementary
sequences are held together by hydrogen bonding between A/T or C/G
nucleotides. Another example is the triple helical structure of collagen.
Under the appropriate thermodynamic conditions the multi-stranded
structure becomes unstable and the strands dissociate from each other.
For a DNA molecule, the dissociation of the two strands is usually
referred to as the melting transition or denaturation; this happens
when the temperature is increased typically above $80^\circ$ C in
standard conditions~\cite{wart85}.  In view of its importance in many
biological and biotechnological processes, various models of DNA melting
have been developed to study the thermodynamical behavior as function
of the sequence length and composition.  Models originally proposed
by Poland and Scheraga \cite{pola66} (PS) and by Peyrard and Bishop
\cite{peyr89} (PB) take into account the different unbinding energies
required to dissociate complementary G/C or A/T pairs using various
types of approximations, yet they are simple enough to deal with very
long DNA sequences.  Using an appropriate choice of parameters they
both fit reasonably well experimental data for the melting temperature
of DNA sequences \cite{camp98}.

The melting {\em dynamics} has also been studied.  In 1986, Baumg\"artner
and Muthukumar~\cite{baumgartner86} performed Monte Carlo simulations
of the disentangling of two initially intertwined chains. After an
initial softening of the original double helix, they reported that
the time required for the actual unraveling of the chains scales
with polymer length as a power law with exponent $3.3 \pm 0.2$.
Computer technology at that time allowed for simulations of 2, 4 and
8 turns only, and their fitted exponent might very well change with
increasing chain length. The dynamics of the opening of bubbles in an
entangled chain has also been studied \cite{PS,mare02,kunz07,baie09}.
In addition different extensions of the PS and PB models were
considered to include helical degrees of freedom of DNA molecules
\cite{cocc99,rudn02,barb03,kaba09}.  These extensions allow for
rapid computations of equilibrium and dynamical properties of the
melting transition, at the cost of introducing approximations which
are poorly controlled and sometimes even questionable.  The moves
introduced to update PS models usually neglect the helicity of dsDNA,
which is locally conserved due to impenetrability of the two strands
(the unbinding of the two strands forming the double helix requires a
release of the twist through a rotation of these strands with respect
to each other). With the denaturation times $\tau_d$ characterized by
the scaling law $\tau_d\sim N^\beta$ for the DNA strand length $N$ (in
base pairs), the resulting values of $\beta$ thus range from $\beta=0$
\cite{kunz07} to $\beta=4/3$ \cite{mare02}.  If local moves preserving
the DNA helicity are introduced, a very slow melting is instead observed,
with $\beta\approx 3$ \cite{baie09}.  A second simplification intrinsic in
PS models is that helical fragments and loops are described by equilibrium
partition functions.  This description can be too simplified in systems
where the dynamics of the unwinding process is too fast to allow for the
full structural relaxation within these loops. PB models have as a major
simplification the possibility for complementary bases to orbit around
a virtual central axis of the macromolecule.  This yields denaturation
times scaling linearly with the chain length~\cite{barb03}.  Given such
spread of results, and uncontrolled simplifications in these models,
we must conclude that we are still missing the understanding of how DNA
length affects the time for its strands to disentangle.

\begin{figure}[!tb]
\includegraphics[height=5cm]{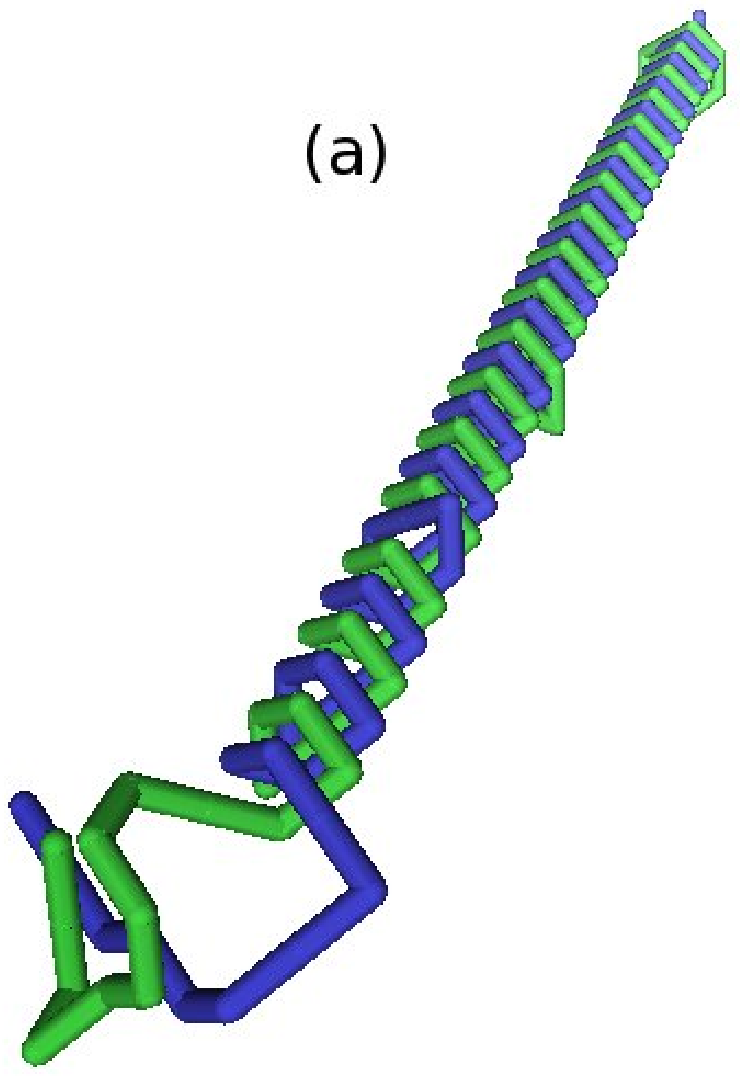}
\includegraphics[height=6cm]{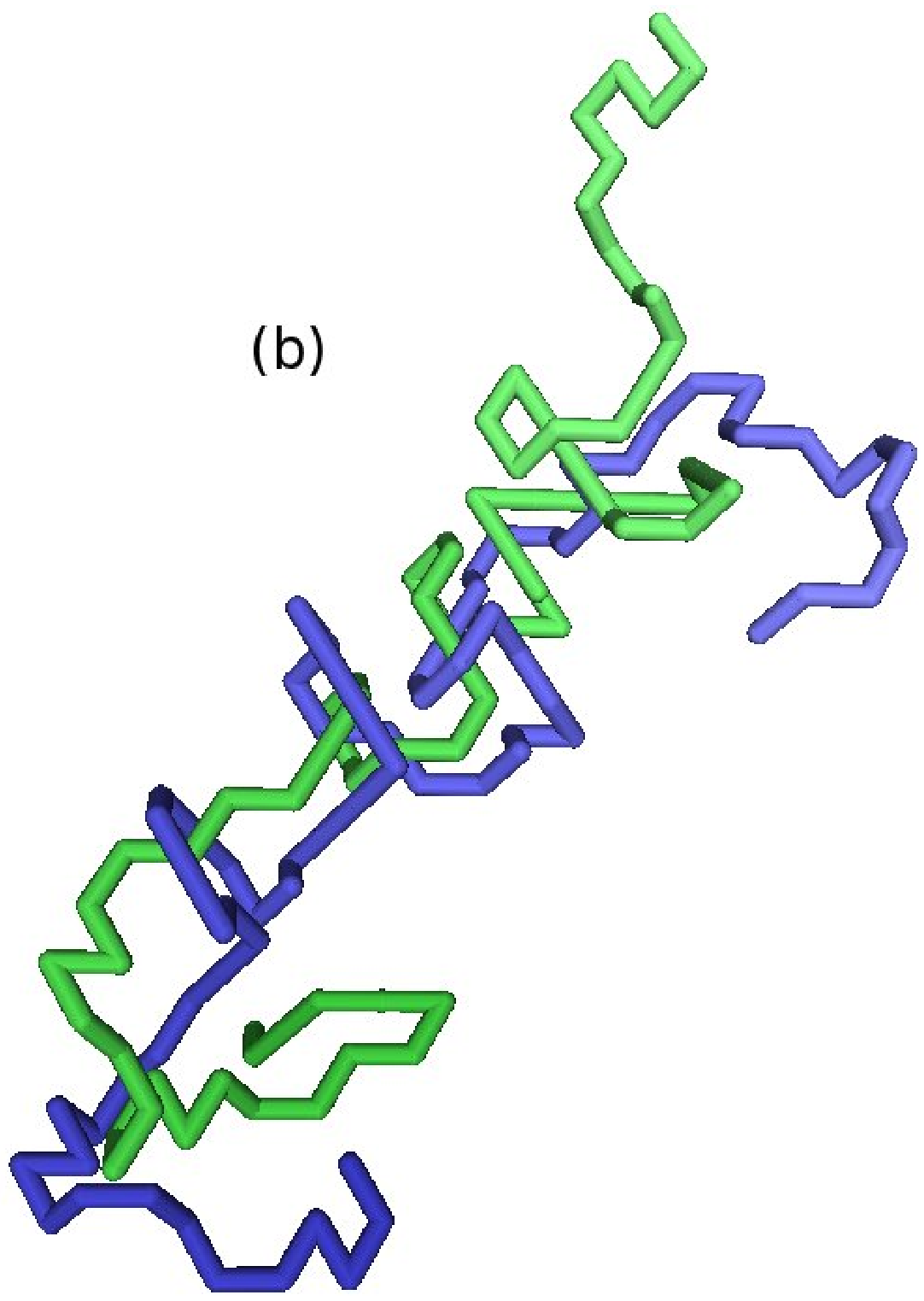}
\includegraphics[height=6cm]{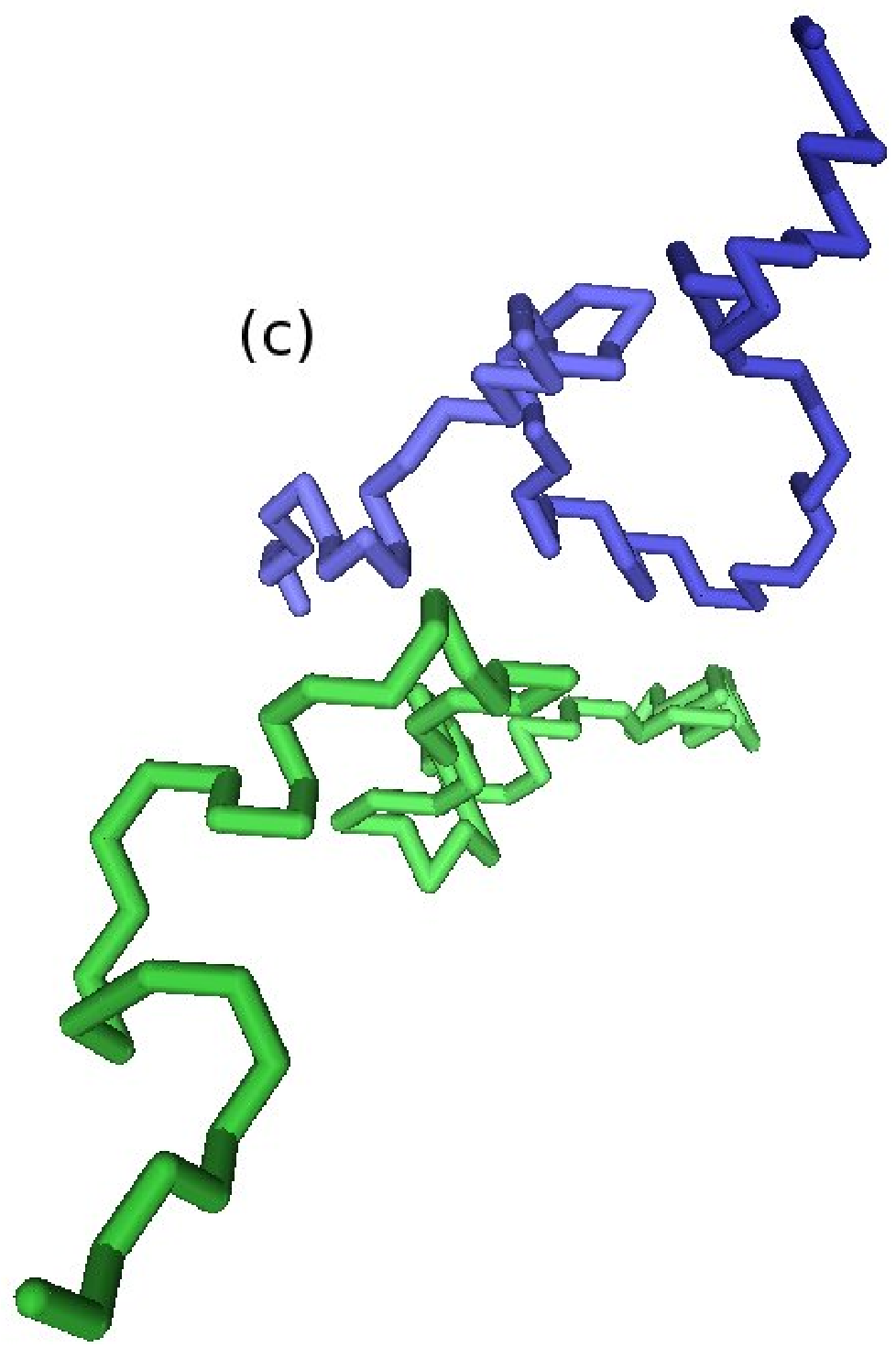}
\caption{(Color online) Snapshots of the polymer configurations during unwinding for
two strands of length $N=100$ each. The initial conformation is fully
double helical all along its length. (a) Snapshot after short time from
the beginning of the simulation; opening begins mainly from the two ends,
although small bubbles within the chain are also visible. (b) Snapshot
at later times. (c) Separation.}
\label{fig01}
\end{figure}

In order to gain more direct insight into the dynamics of the melting
process of double-stranded molecules we investigate the unwinding
dynamics of double-stranded three-dimensional long polymers using Monte
Carlo simulations in the absence of hydrodynamic interactions. No
binding energies between the two strands are taken into account during
the unwinding process, corresponding to the case of a double helix
brought rapidly to a temperature well above its melting point. For
such a setup, the strands unwind from each other as it is entropically
favorable for them to do so. We follow a procedure very similar to that
of Ref.~\cite{baumgartner86}, except that our chains are much longer,
and more tightly wound. Figure~\ref{fig01} shows three configurations:
(a) at the early stages of unwinding, (b) during the unwinding process,
and (c) at the end of the unwinding, when the two stands are separating
from each other.  The main scope of this manuscript is two-fold:
in section~\ref{sec:time} we study the unwinding time $\tau_u$ as a
function of $N$ --- high precision simulations for polymers of length
up to $N=1000$ show that $\tau_u$ scales as a power-law $\tau_u\sim
N^\beta$ with $\beta = 2.57\pm0.03$. In section~\ref{sec:size} we show
that the intermediate conformations can be characterized as a tightly
wound inner part, to which unwound single strands are connected. The
length of the unwound strands increases with time in a powerlaw fashion
as $l(t)\sim t^{0.39}$.  With a theoretical argument, we find an upper
bound for the radius of gyration of the loose ends, which excludes the
equilibrium value. This demonstrates that the unwinding is a far from
equilibrium process.

\section{Unwinding time}
\label{sec:time}

In the simulation the polymers reside on a face-centered-cubic
lattice with a lattice spacing of $\sqrt{2}$, and are initialized in a
double-helical state. The polymers evolve in time through a long sequence
of single-monomer moves, under the restriction that at all times,
the polymer backbones are self- and mutually avoiding. Each allowed
move occurs with a statistical rate of unity. To give the polymers some
elasticity, the self-avoidance condition is lifted for monomers which are
direct neighbors along the same chain.  A detailed description of this
lattice polymer model, its computationally efficient implementation,
and a study of some of its properties and applications can be found
in \cite{heuk03}.  This model reproduces known features of the Rouse
dynamics \cite{panja} and of the equilibrium properties \cite{baie10}
of single self-avoiding polymers. As the moves respect the no-crossing
condition between strands, we expect that the long time behavior of
unwinding discussed in this study is of universal nature and is not
affected by microscopic details and lattice effects.

\begin{figure}[!tb]
\includegraphics[width=12cm]{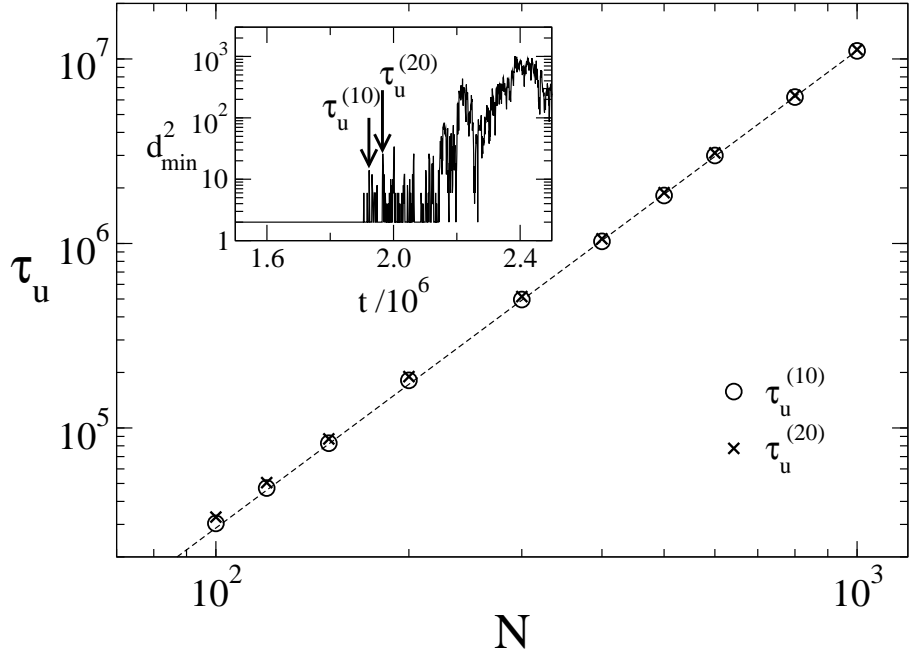}
\caption{
Double-logarithmic plot of the average unwinding time as a function of
strand length. The circles are obtained with an unwinding threshold of
$d_0^2=10$, while the crosses correspond to a threshold of $d_0^2=20$. The
straight dashed line is a fit to the data corresponding to an unwinding
exponent of $\beta=2.57$. Inset: Plot of $d^2_{\rm min}(t)$ vs. time
for a run with strands of length $N=500$.  The arrows indicate the first
time that this distance reaches its threshold value $d_0^2=10$ or 20.
\label{fig02}}
\end{figure}

Let $\vec{r}_i^{(1)}(t)$ and $\vec{r}_i^{(2)}(t)$ be the lattice positions
of the $i$-th monomers on the two strands at time $t$ ($0 \leq i
\leq N$).  We consider the minimal distance between two strands defined
as $\displaystyle{d_{\min} (t) = \min_{i,j} \left| \vec{r}_i^{(1)}(t)
- \vec{r}_j^{(2)}(t) \right|}$. The inset in Figure \ref{fig02} shows
a plot of $d^2_{\rm min} (t)$ as a function of time for a run. The
choice of an initial double-helical conformation implies that $d_{\rm
min}=\sqrt{2}$ at $t=0$. Note that $d_{\rm min} (t)$ remains constant up
to a time $t \approx 1.9 \cdot 10^6$ in the inset of Fig.~\ref{fig02}
and then starts fluctuating and increasing in time.  We define the
unwinding time $\tau_u$ as the time at which $d_{\rm min} (t)$ exceeds
some threshold value for the first time.  For the threshold value $d_0$
we took $d_0^2 =10$ and $d_0^2= 20$. The higher threshold value gives
a slightly higher estimate of the unwinding time [$\tau_u^{(10)}$ and
$\tau_u^{(20)}$ in Fig.~\ref{fig02}].  However, as the polymer length
increases the ratio of $\tau_u^{(10)}/\tau_u^{(20)}$ converges to $1$
as shown in Table~\ref{tab}, hence the two quantities have the same
scaling behavior in $N$.

Figure~\ref{fig02} and Table \ref{tab} show the behavior of $\tau_u^{(10)}$
and $\tau_u^{(20)}$ as a function of polymer length. We note that the scaling of 
unwinding times is a power of the strand length: From a linear regression of
the data for $N > 30$ we find the values $\beta= 2.58 \pm 0.03$
($d_0 = 10$) and $\beta = 2.56 \pm 0.03$ ($d_0 = 20$), from which we obtain
the result anticipated above $\beta = 2.57 \pm 0.03$.

\section{Characterization of intermediate conformations}
\label{sec:size}

\begin{figure}[!tb]
\includegraphics[width=9cm]{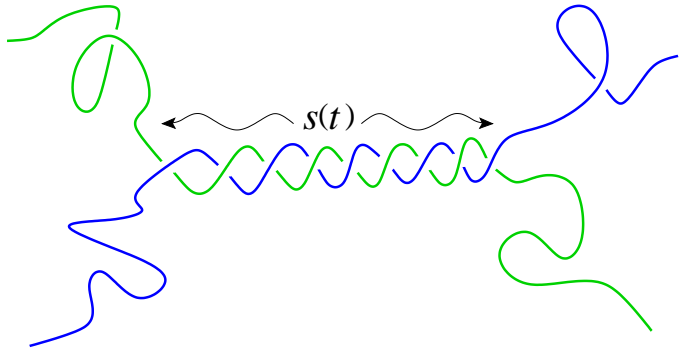}
\caption{Sketch of a double-stranded polymer during the
unwinding dynamics. At time $t$ we expect to find a double stranded
region of curvilinear length $s(t)$ terminates with two single strands
of lengths $(N-s(t))/2$ at both edges, $s(t)$ being a decreasing
function of $t$.}
\label{fig03}
\end{figure}

\begin{figure}[!bt]
\includegraphics[width=12cm]{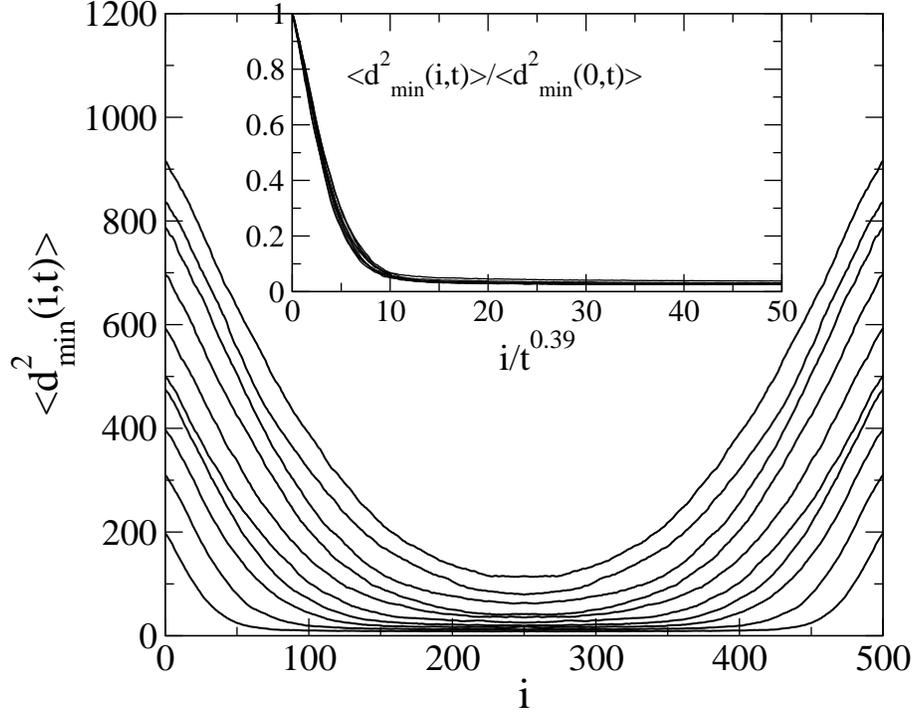}
\caption{
Average squared minimal distance  $\langle d^2_{\min} (i,t) \rangle$ as
a function of monomer number $i$, for polymers with length $N=500$. From
bottom to top, the curves are obtained at times $t/10^5 = 1, 2 \ldots 10$.
Inset: Collapse of $\langle d^2_{\min} (i,t) \rangle/\langle d^2_{\min}
(0,t) \rangle$ for polymers of length $N=1000$ again at times $t/10^5 =
1, 2 \ldots 10$.  The vertical scale is normalized so that all curves
start from a common value at $i=0$. The horizontal scale is divided
by $t^{0.39}$.  This exponent is consistent with that obtained from the
scaling of the unwinding time.
}
\label{fig04}
\end{figure}

Given the topological constraint each strand faces while unwinding,
we expect the unwinding dynamics to unroll from the two ends of the
initial double-stranded complex, progressing inwards with increasing
time (Fig.~\ref{fig03}).  Note that because of the elasticity of the
model used in the simulations, a partial opening up of the inner wound
part is not ruled out by the model: it is the physics of the problem
that seems to suppress this.  In order to connect this physical picture
with the observed scaling $\tau_u\sim N^{2.57\pm0.03}$ we considered the
quantity $\displaystyle{d_{\min}(i,t) = \min_j \left|\vec{r}_i^{(1)}(t)
- \vec{r}_j^{(2)}(t) \right|}$ which is the minimal distance from the
$i$-th monomer of the first strand to any other monomer of the second
strand. Its average square $\langle d^2_{\min} (i,t) \rangle$ is plotted
in Fig.~\ref{fig04} for a strand length equal to $N=500$. The different
data are for increasing time step snapshots taken at time intervals
equal to $t/10^5 = 1, 2, \ldots, 10$ (from bottom to top). The quantity
$\langle d^2_{\min} (i,t) \rangle$ is minimal in the middle, while it
increases in time from the two edges, in agreement with the physical
picture proposed in Fig.~\ref{fig03}.  We consider now the normalized
profile $\langle d^2_{\min}(i,t)\rangle/\langle d^2_{\min}(0,t) \rangle$.
For this quantity we expect the following scaling behavior, as a function
of the distance from the end monomer ($i=0$)
\begin{equation}
\frac{\langle d^2_{\min}(i,t)\rangle}{\langle
d^2_{\min}(0,t) \rangle} = f(i/l(t))
\end{equation}
with $f()$ a scaling function and $l(t)$ a characteristic length depending
on time. The inset of Fig.~\ref{fig04} show that the normalized profiles
at different times collapse when a rescaling $i/t^{0.39}$ is used, which
implies $l \sim t^{0.39}$. This is consistent with the exponent determined
from the scaling of unwinding time as $l \sim t^{1/\beta} = t^{0.39}$.

In order to gain insight into the unwinding we set up a simple analytical
model of the process, assuming that the unwinding is sufficiently slow
so that the conformation of the loose strands can be approximated
by equilibrium ones at all times. We will show that this approach
predicts an unwinding dynamics which is slower than what is observed
in simulations.  We therefore conclude that the unwinding we observe is
a far from equilibrium process.

Consider an intermediate conformation which consists of two
single-stranded coils of $N-s(t)$ attached to a double stranded helical
part of length $s(t)$. We estimate the free energy $F$ as a function
of $s(t)$ from the partition function for a double helical segment of
length $s$ is $Z_{\rm helix} \sim \mu_h^s$ and that of a single stranded
coil $Z_{\rm coil} \sim \mu_c^{N-s}$, where $\mu$ are the connectivity
constants. In other words
\begin{equation} 
\beta F = - s \log(\mu_h)+ 2(N-s) \log(\mu_c).
\label{energy_balance2} 
\end{equation} 
As an infinitesimal portion $ds$ of the double helix unwinds, the change
in free energy is thus given by
\begin{equation} 
dF = -\beta^{-1} \log(\mu_c^2/\mu_h)ds \equiv -K_1 ds.  
\end{equation} 
During this infinitesimal unwinding, the single-stranded coils are
displaced over a distance $dr$ due to the rotational motion around the
axis of the helix: the coils describe a fraction of a circle of radius
$R_v$ perpendicular to the axis of the helix, where $R_v$ is the
distance from the helix axis to the coil's center of mass. In equilibrium
we expect $R_v \sim (N-s)^\nu$, with $\nu$ the Flory exponent. This
implies that
\begin{equation} 
dr \propto  R_v ds \propto (N-s)^\nu ds.  
\label{rands} 
\end{equation} 
During this process the work done against friction equals 
\begin{equation} 
dW = \gamma \dot{r} dr = \gamma R_v^2 \dot{s}~ds, 
\label{energy_balance} 
\end{equation}
where for Rouse dynamics the friction on the single-stranded coils is
proportional to their lengths: $\gamma \propto N-s$.  Since the work done
against friction cannot exceed the available free energy, we obtain the
inequality \begin{equation} K_2 (N-s)^{1+2\nu} \dot{s}~ds \le K_1 ds.
\end{equation} If this inequality were saturated, the unwinding process
would be described by 
\begin{equation} 
K_1/K_2  =-  [N-s(t)]^{2\nu + 1} \frac{ds}{dt}.  
\label{diff_eq} 
\end{equation} 
The integration of Eq.~(\ref{diff_eq}) yields a scaling of the unwinding
time as $\tau_u \sim N^{2\nu + 2} = N^{3.18}$ (since $\nu = 0.59$),
which is obtained from Eq.~(\ref{diff_eq}) by setting $s(\tau_u)=N$.
Indeed, a visual investigation of the unwound parts during simulations
reveals that these show a tendency to be spiral-like, and contain a
significant amount of coiling.  A precise quantification of the amount
of spiraling is very difficult due to the lack of a clean definition
of the central axis of the spiral.  This argument predicts a very slow
unwinding compared to that which is actually observed in simulations,
suggesting that unwinding proceeds through non-equilibrium states.
In an early study of unwinding \cite{baumgartner86} the scaling $\tau
\sim N^{3.3(2)}$ was computed for shorter polymers (up to $N = 65$;
with 2, 4 or 8 turns in the double helix) and for helices less tightly
bound than considered here. The exponent of~\cite{baumgartner86} most
likely describes a pre-asymptotic scaling regime.  The exponent reported
in \cite{baumgartner86} is however compatible with that predicted from
Eq.~(\ref{diff_eq}). This suggests that the early stages of unwinding
are probably well captured by the mechanisms underlying the derivation of
Eq.~(\ref{diff_eq}), but that the asymptotic scaling regime is dominated
by a mechanism which is faster for longer chains.

\section{Summary and conclusions}

Summarizing, we introduced a lattice model for studying the unwinding
dynamics of a long three-dimensional double stranded polymer, with
excluded volume effects taken into account.  The lattice nature
of the model, combined with an efficient encoding of the dynamics,
allows one to simulate long polymers (up to $N=10^3$) for very long
time ($t = 10^7$ Monte Carlo steps).  Our numerical results show that
the unwinding time scales with the polymer length as a power-law with
exponent $\beta=2.57(3)$ with Rouse dynamics.  An analysis of a simple
analytical model of the process suggests that the unwinding we observe
is a far from equilibrium process, therefore it cannot be understood
in terms of a slow dynamics evolving through quasi-equilibrium states,
e.g. using free energy arguments.

\vfill\eject
\newpage

\begin{table}[t]
\begin{tabular}{c|c|c|c}
$N$ & $\tau_u^{(10)}$ & $\tau_u^{(20)}$ & $\tau_u^{(20)}/\tau_u^{(10)}$\cr
\hline
60   & $1.23\cdot 10^4$ & $1.12\cdot 10^4$ & 1.093 \cr
80   & $1.80\cdot 10^4$ & $1.97\cdot 10^4$ & 1.096 \cr
100  & $3.04\cdot 10^4$ & $3.29\cdot 10^4$ & 1.082 \cr
120  & $4.73\cdot 10^4$ & $5.06\cdot 10^4$ & 1.070 \cr
150  & $8.27\cdot 10^4$ & $8.75\cdot 10^4$ & 1.055 \cr
200  & $1.81\cdot 10^5$ & $1.90\cdot 10^5$ & 1.048 \cr
300  & $4.97\cdot 10^5$ & $5.16\cdot 10^5$ & 1.038 \cr
400  & $1.03\cdot 10^6$ & $1.06\cdot 10^6$ & 1.032 \cr
500  & $1.82\cdot 10^6$ & $1.88\cdot 10^6$ & 1.036 \cr
600  & $3.00\cdot 10^6$ & $3.10\cdot 10^6$ & 1.032 \cr
800  & $6.22\cdot 10^6$ & $6.35\cdot 10^6$ & 1.021 \cr
1000 & $1.11\cdot 10^7$ & $1.13\cdot 10^7$ & 1.017
\end{tabular}
\caption{Average unwinding times $\tau_u^{(10)}$ and $\tau_u^{(20)}$ as a
function of polymer length $N$ for threshold values $d_0^2 = 10$
and $d_0^2= 20$, and the ratio of these two times. 
Times are obtained by averaging over 120 simulations.
}
\label{tab}
\end{table}
\vfill\eject
\newpage

\end{document}